\documentstyle[12pt,amssymb,epsfig]{article}
\renewcommand{\theequation}{\arabic{section}.\arabic{equation}}
\def \thesection {\arabic{section}.}

\newcommand{\be}{\begin{equation}}
\newcommand{\ee}{\end{equation}}
\newcommand{\ba}{\begin{eqnarray}}
\newcommand{\ea}{\end{eqnarray}}
\newcommand{\baa}{\begin{eqnarray*}}
\newcommand{\btab}{\begin{tabular}}
\newcommand{\etab}{\end{tabular}}
\newcommand{\eaa}{\end{eqnarray*}}
\def \labeltest #1 {\label{#1}}
\newcommand\re[1]{(\ref{#1})}
\def \qqquad {\qquad\quad}

\newcommand\lr[1]{{\left({#1}\right)}}

\def \matrix #1 {\left(\begin{array}{cc} #1 \end{array}\right)}

\newcommand \vev [1] {\langle{#1}\rangle}
\newcommand \VEV [1] {\left\langle{#1}\right\rangle}

\newcommand \ket [1] {|{#1}\rangle}

\def \e {\mbox{e}}

\def\II{\hbox{{1}\kern-.25em\hbox{l}}}

\catcode`\@=11
\def\numberbysection{\@addtoreset{equation}{section}
                     \def\theequation{\thesection\arabic{equation}}}
\numberbysection

\begin{document}

\begin{titlepage}
\begin{flushright}
\begin{tabular}{l}
LPT--Orsay--00--63\\
hep-ph/0007005
\end{tabular}
\end{flushright}

\vskip3cm

\begin{center}
  {\large \bf
  On power corrections to the event shape distributions in QCD}

\vspace*{1cm}
{\sc G.P.~Korchemsky} and {\sc S.~Tafat}

\vspace*{0.1cm}
{\it
Laboratoire de Physique Th\'eorique%
\def\thefootnote{\fnsymbol{footnote}}%
\footnote{Unite Mixte de Recherche du CNRS (UMR 8627)},
Universit\'e de Paris XI, \\
91405 Orsay C\'edex, France
                       }

\vskip0.8cm
{\bf Abstract:\\[10pt]} \parbox[t]{\textwidth}
{We study power corrections to the differential thrust, heavy jet mass and
$C-$parameter distributions in the two-jet kinematical region. We argue that
away from the end-point region, $e\gg \Lambda_{\rm QCD}/Q$, the leading
$1/Q-$power corrections are parameterized by a single nonperturbative scale
while for $e\sim\Lambda_{\rm QCD}/Q$ one encounters a novel regime in which
power corrections of the form $1/(Qe)^n$ have to be taken into account for
arbitrary $n$. These nonperturbative corrections can be resummed and factor out
into a universal nonperturbative distribution, the shape function, and the
differential event shape distributions are given by convolution of the shape
function with perturbative cross-sections. Choosing a simple ansatz for the
shape function we demonstrate a good agreement of the obtained QCD predictions
for the distributions and their lowest moments with the existing data over a
wide energy interval.}

\vskip1cm

\end{center}

\end{titlepage}

\newpage



\section{Introduction}

Analysis of hadronization effects to the final states in $\e^+\e^--$annihilation
has became the subject of active QCD studies \cite{Exp}. There exist infrared and
collinear safe event shape variables for which perturbative QCD can be applied
at large center-of-mass energies $s=Q^2$ to calculate their differential
distributions and mean values as series in $\alpha_s(Q)$. It has been observed
many years ago \cite{Old} that for some shape variables like thrust, $t$, and
heavy jet mass, $\rho$, perturbative QCD predictions deviate from the data by
corrections suppressed by powers of the large energy scale $1/Q^p$, with the
exponent $p$ depending on the variable and $p=1$ for $t-$ and $\rho-$variables.
Such hadronization corrections were measured experimentally \cite{Exp} over a
wide energy interval $14\le\sqrt{s}/{\rm GeV}\le 189$ and were found to have a
different form for the differential event shape distributions, ${d\sigma}/{d
e}$, as compared to their mean values, $\vev{e} =\sigma_{\rm tot}^{-1}\int
de\,e{d\sigma}/{d e}$. For the mean value $\vev{e}$ the leading power correction
is parameterized by a nonperturbative {\it scale\/} $\lambda_p$ of dimension
$p$, while hadronization corrections to the differential distribution are
described by a {\it function} $f_{\rm hadr}(Q,e)$ depending on both the shape
variable and the center-of-mass energy
\be
\vev{e} = \vev{e}_{_{\rm PT}} + \lambda_p/Q^p\,,\qquad
\frac1{\sigma_{\rm tot}}\frac{d\sigma}{d e} =
\frac{d\sigma_{_{\rm PT}}}{d e} + f_{\rm hadr}(Q,e)
\label{f-hadr}
\ee
with $e$ denoting a general event shape variable $(e=t\,,\rho\,,C\,,...)$ and the
subscript PT referring to perturbative contribution, $\vev{e}_{_{\rm PT}}=\int
de\,e{d\sigma_{_{\rm PT}}}/{d e}$. Obviously, the hadronization corrections to
the differential distributions have a richer structure then those to the mean
values. For instance, nonperturbative scales $\lambda_p$ parameterizing power
corrections to $\vev{e}$ are defined by the moment $\int de\,e f_{\rm
hadr}(Q,e)$.

Power corrections in \re{f-hadr} are associated with hadronization effects in
${\rm e}^+{\rm e}^--$final states and, as a consequence, the magnitude of the
scales $\lambda_p$ and the function $f_{\rm hadr}(Q,e)$ cannot be calculated
within perturbative QCD approach. However it was recognized some time ago
\cite{W,KS,AZ,BB}, that analysis of infrared renormalon ambiguities of perturbative QCD
series suggests the value of dimensionless exponents $p$ as well as the
dependence of the function $f_{\rm hadr}(Q,e)$ on the large scale $Q$. Namely,
perturbative QCD series generate power corrections of the form \re{f-hadr}
through IR renormalons contribution but fail to predict uniquely their values --
it is only the sum of perturbative and nonperturbative contributions that
becomes well-defined \cite{Beneke}. To give a meaning to the perturbative series
in \re{f-hadr} one has to regularize IR renormalon singularities. This can be
done in two different ways: one can specify a particular prescription for
integrating IR renormalon singularities like principal value prescription
\cite{GG}. Alternatively, one can avoid IR renormalon ambiguities by introducing
an explicit IR cut-off $\mu$ on momenta of soft particles in perturbative
expressions. In this case, one can either impose a ``hard'' IR cut-off on
momenta of soft particles in the Feynman integrals, $k_\perp >\mu$, \cite{KS} or
replace QCD coupling constant by a effective IR finite coupling constant which
coincides with $\alpha_s(k_\perp)$ at large scale $k_\perp$ and deviates from it
at $k_\perp<\mu$ \cite{W,DMW}. Following each of these ways, one specifies
perturbative ($\mu-$dependent) contribution to \re{f-hadr} including
perturbatively induced power corrections. Still, there exists a genuine
nonperturbative contribution to the event shapes coming from the QCD dynamics at
scales below $\mu$. This contribution cannot be determined from the analysis of
perturbative QCD series while its magnitude depends on the choice of the IR
regularization and, as a consequence, on the IR cut-off $\mu$.

For some hadronic observables like mean values of the event shapes and their
differential distributions away from the end-point region, leading
nonperturbative power corrections can be parameterized using different IR
renormalon inspired phenomenological models \cite{W,AZ,KS,Beneke}. Their
predictions agree well with the experimental data and the extracted values of
phenomenological nonperturbative parameters exhibit approximate universality.
Despite a phenomenological success of these models, it remains still unclear what
is the physical meaning of new nonperturbative QCD scales and what is the origin
of the universality property within QCD. In the present paper we address these
problems using the factorization properties of the event shape distributions
established in \cite{KS-shape}. We shall argue that nonperturbative power
corrections to the thrust, heavy jet mass and $C-$parameter distributions are
described by the universal shape function which is a new nonperturbative QCD
distribution measuring the energy flow in the two-jet final states in
$\e^+\e^--$annihilation.

The paper is organized as follows. In Sect.~2 we discuss the general properties
of power corrections to the event shape distributions in the end-point region. In
Sect.~3 we formulate the factorization procedure and define the shape function.
In Sect.~4 we show that the differential event shape distributions are given by
the convolution of the resummed perturbative cross-sections with universal shape
function. Choosing a simple ansatz for this function we compare QCD predictions
with the existing data. In Sect.~5 we apply the obtained expressions to calculate
the power corrections to the first two moments of the distributions. Concluding
remarks are given in Sect.~6.

\section{Event shape distributions}

In this paper we shall consider three event shape variables: thrust $T$, heavy
jet mass $\rho$ and $C-$parameter. They are defined in the standard way as
\cite{KNMW}
\be
T=\max_{\vec n_T} \frac{\sum_k |{\vec p_k\cdot\vec n_T}|}{\sum_k |{\vec
p_k}|}\,,\qqquad
\rho = \max\lr{\frac{M_R^2}{Q^2},\frac{M_L^2}{Q^2}}\,,
\ee
where $M_R^2$ and $M_L^2$ denote the total invariant masses flowing into the
right and left hemispheres with respect to the plane orthogonal to the thrust
axis $\vec n_T$. The $C-$parameter is given by
\be
C=3(\theta_1\theta_2+\theta_2\theta_3+\theta_3\theta_1)
\ee
with  $\theta_j$ being eigenvalues of space-like part of the energy-momentum
tensor $\Theta^{\alpha\beta}=\sum_k {p_k^\alpha p_k^\beta}/|p_k|/\sum_j |p_j|$.

Introducing the new variable $t=1-T$ one notices that thus defined event shapes
$e=(t\,,\rho\,,C)$ have a number of common features. Lowest order perturbative
QCD calculation leads in all three cases to the following expression for the
differential distribution for $e>0$
\cite{KNMW}
\be
\frac{d\sigma_{_{\rm PT}}}{d e}= \frac{\alpha_s(Q)}{2\pi} A_e(e)\theta(e_{\rm
max}-e) +\lr{\frac{\alpha_s(Q)}{2\pi}}^2 B_e(e)+{\cal O}(\alpha_s^3)\,,
\label{A_e}
\ee
where $A_e$ and $B_e$ are known coefficient functions and normalization is chosen
as $\int de\frac{d\sigma_{_{\rm PT}}}{d e}=1$. Lowest order correction $A_e$ gets
contribution only from the three-particle final state which populates the
kinematic region $0\le e \le e_{\rm max}$ with $t_{\rm max}=\rho_{\rm max}=1/3$
and $C_{\rm max}=3/4$. Away from the end-point region, $e\gg \Lambda_{_{\rm
QCD}}/Q$, the perturbative expansion \re{A_e} is well-defined and it describes
the final states consisting of particles with relative transverse momentum that
scales at large center-of-mass energy as $\sim Q$. As $e$ approaches the
three-particle upper limit, $e=e_{\rm max}$, $A_t$ and $A_\rho$ vanish while
$A_C$ takes a finite value
\cite{KNMW,ERT}
\ba
&&
A_t(1/3) = A_\rho(1/3) =0 \,,
\nonumber
\\
&& A_C(C) = \frac{256}{243}\pi\sqrt{3} C_F\left[ 1-\frac83\lr{C-\frac34} + {\cal
O}\lr{(C-3/4)^2}\right]\,.
\label{A's}
\ea

For $e=(t\,,\rho\,,C) \to \Lambda_{_{\rm QCD}}/Q$ the final states consist of two
narrow jets with invariant mass $M^2_{R,L}\sim
\Lambda_{_{\rm QCD}}Q$. Examining \re{A_e} one finds that $A_e$
diverges in the end-point region $e\to 0$ as \cite{KNMW,ERT,CTTW,CW}
\be
A_e(e) = \frac{4C_F}{e}\left[\ln{\frac{e_0}{e}}-\frac34\right] + {\cal O}(\ln e)
\ee
with $t_0=\rho_0=1$ and $C_0=6$. Similar Sudakov-type corrections appear to
higher orders, $\alpha_s^N\ln^{2N-n}e/e$ with $n\ge 0$, and need to be resummed
\cite{CTTW,CW}. They originate from the effects of collinear splitting of quarks and gluons
inside two narrow energetic jets and their interaction with surrounding cloud of
soft gluons. The underlying QCD dynamics depends on two infrared scales, $Q e$
and $Q^2 e$, such that $1/Q\ll 1/(Qe^{1/2})\ll 1/(Q e)$. The smallest scale $Q e$
sets up the typical energy carried by soft gluons, while the scale $Q\sqrt{e}$
defines the transverse momenta of the jets, $k_\perp^2= Q^2 e$. Applying the
standard IR renormalon analysis and examining sensitivity of perturbative
expressions with respect to emission of particles on each of these scales, that
is soft gluons with energy $\sim Q e$ and collinear particles with the
transverse momentum $\sim Q^2e$, one finds that nonperturbative corrections to
the differential distribution appear suppressed by powers of both scales. Then,
in the end-point region, $e={\cal O}(\Lambda_{_{\rm QCD}}/Q)$, we may use the
fact that $Q e={\cal O}(\Lambda_{_{\rm QCD}})$ and expand the differential
distribution in powers of larger scale $Q^2 e$. Keeping only the leading term of
the expansion one gets
\be
\frac1{\sigma_{\rm tot}}\frac{d\sigma}{d e} = \sigma_0\lr{\alpha_s(Q),\ln e, \frac1{Qe}}
+ {\cal O}\lr{\frac1{Q^2e}} \,,
\label{sigma_0}
\ee
where $\sigma_0$ resums perturbative corrections in $\alpha_s(Q)$ as well as
power corrections on the smallest scale $Qe$
\be
\sigma_0=
\frac{d\sigma_{_{\rm PT}}}{d e}
+\sum_{k=1}^\infty \frac{\lambda_k}{(Q e)^k}\Sigma_{k}(\alpha_s(Q),\ln e)
\,.
\label{gen-exp}
\ee
Here, $\Sigma_k$ are dimensionless perturbative coefficient functions and
$\lambda_k$ are some nonperturbative scales, depending, in general, on the
choice of the event shape variable. Using \re{gen-exp} we notice that the power
corrections have a different form for $e\gg\Lambda_{_{\rm QCD}}/Q$ and $e
\sim\Lambda_{_{\rm QCD}}/Q$.

For $e$ away from the end-point region, $e\gg
\Lambda_{_{\rm QCD}}/Q$, one may keep in
\re{gen-exp} only the first term
\be
\frac1{\sigma_{\rm tot}}\frac{d\sigma}{d e}
=\frac{d\sigma_{_{\rm PT}}}{d e} +\frac{\lambda_1}{Q e}\Sigma_{1}(\alpha_s(Q),\ln
e) + {\cal O}\lr{\frac1{(Qe)^2}}\,.
\label{off-peak}
\ee
The coefficient function $\Sigma_{1}$ can be found using the well-known property
\cite{KS,DW} that the {\it leading\/} $1/Q-$power correction to the differential
distribution \re{off-peak} is generated by a shift of perturbative spectrum,
$e\to e-\lambda_1/Q$. This leads to
\be
\Sigma_{1}(\alpha_s(Q),\ln e)=-\frac{d}{d\ln e}\left[
\frac{d\sigma_{_{\rm PT}}}{d e}\right]\,.
\label{Sigma1}
\ee
Then, it follows from \re{off-peak} that for $e\gg\Lambda_{_{\rm QCD}}/Q$ the
leading power corrections to the differential distributions have a rather simple
structure: they are parameterized by a single nonperturbative scale $\lambda_1$.
The same scale determines $1/Q-$power correction to the mean value $\vev{e}$. The
QCD predictions \re{off-peak} are in a good agreement with the experimental data
and the value of $\lambda_1$ has been fitted for different shape variables
\cite{Exp}. It is worthwhile to note that in the performed analysis of power
corrections to the differential distributions \cite{Exp} the fitting range of
event shape variables was restricted to the region $e\gg\Lambda_{_{\rm QCD}}/Q$.
At the same time, as we shall argue below, it is in the region
$e\sim\Lambda_{_{\rm QCD}}/Q$ where a novel QCD regime is realized and the
structure of hadronization corrections is drastically changed.

For $e \sim \Lambda_{_{\rm QCD}}/Q$ one finds that all terms in \re{gen-exp}
become equally important and, therefore, need to be resummed to all orders in
$1/(Qe)$. The resummation is based on the remarkable factorization properties of
the differential distributions. As was shown in \cite{KS-shape}, the
nonperturbative corrections to the leading asymptotic term $\sigma_0$ are
factorized out into nonperturbative distribution function, the so-called shape
function. The general factorized expression for differential distribution looks
like \cite{KS-shape}
\be
\frac1{\sigma_{\rm tot}}\frac{d\sigma}{d e}=
\int_0^{eQ} d\varepsilon f_e(\varepsilon)\frac{d\sigma_{_{\rm PT}}
(e-\frac{\varepsilon}{Q})}{d e} +{\cal O}\lr{\frac1{Q^2e}} \,.
\label{factor}
\ee
Its explicit form depends on the choice of the event shape variable
$e=(t\,,\rho\,,C)$ and will be given in Sect.~4 (see Eqs.~\re{dis-t}, \re{dis-C}
and \re{dis-rho}).  For $e\gg \Lambda_{\rm QCD}/Q$ one can expand the r.h.s.\ of
\re{factor} in powers of $1/Q$ to reproduce the expansion \re{gen-exp} with
\be
\lambda_n=\int d\varepsilon\,\varepsilon^n f(\varepsilon)\,,\qquad
\Sigma_n(\alpha_s(Q),\ln e)=\frac{(-e)^n}{n!}\frac{d^n}{de^n}\left[\frac{d\sigma_{_{\rm PT}}}{de}
\right]\,.
\label{lambda_n}
\ee
Expression \re{factor} has a simple physical interpretation -- nonperturbative
corrections increase invariant masses of jets and effectively shift perturbative
spectrum towards larger values of the shape variables with the weight given by
nonperturbative distribution $f_e(\varepsilon)$.

\section{Factorization and Shape functions}

Factorization relations \re{factor} take a simple form for the radiator functions
$R(e)$ defined as \cite{CTTW}
\be
R(e) = \int_0^e de' \frac1{\sigma_{\rm tot}}\frac{d\sigma}{d e'} \equiv
\vev{\theta(e - e(N))}\,.
\ee
Here, $\vev{...}$ denotes averaging over all possible final states in ${\rm
e}^+{\rm e}^--$annihilation with the weight given by the differential
distribution $1/{\sigma_{\rm tot}}{d\sigma}/{d e}$ and $e(N)$ denotes the value
of the event shape variable $e$ for a given final state $\ket{N}$. Calculating
$R(e)$ in perturbation theory one finds
\be
R_{_{\rm PT}}(e) = 1 - \frac{\alpha_s(Q)}{2\pi}\int_{e}^{e_{\rm max}} de' A_e(e')
+ {\cal O}(\alpha_s^2(Q))\,.
\label{R_pt}
\ee

Close to the two-jet region, $e\to 0$, perturbative expressions for $R(e)$
involve Sudakov logs $\alpha_s^N \ln^{2N-n} e$ with $n\ge 0$. In the case of
event shapes under consideration, $e=(t\,,\rho\,,C)$, these corrections can be
systematically resummed to next-to-leading logarithmic (NLL) order and matched
into exact two-loop perturbative expressions \re{R_pt} \cite{CTTW,CW}. To this
accuracy the weights $e(N)$ can be expressed in terms of the total invariant
masses $M_R^2$ and $M_L^2$ of two jets flowing into the right and left
hemispheres, respectively. Moreover, the $t-$ and $C-$parameters depend only on
the sum of two masses and the corresponding perturbative radiation functions can
be expressed to the NLL approximation as
\cite{CTTW,CW}
\ba
R_t^{_{\rm PT}}(e) &=& \vev{\theta(e - t(N))}_{_{\rm PT}} = \VEV{\theta\left(e -
\frac{M_R^2+M_L^2}{Q^2}\right)}_{_{\rm PT}}
\label{R-t}
\\
R_C^{_{\rm PT}}(e) &=& \vev{\theta(e - C(N))}_{_{\rm PT}} = \VEV{\theta\left(e -
6\frac{M_R^2+M_L^2}{Q^2}\right)}_{_{\rm PT}} = R_t^{_{\rm PT}}(e/6)\,,
\nonumber
\ea
where the subscript PT indicates that the final states in $\e^+\e^--$annihilation
are generating by perturbative branching of outgoing quark and antiquark. The
radiator function for the $\rho-$parameter depends separately on the masses of
two jets. Taking into account that perturbative evolution of two jets is
independent on each other to the NLL approximation one gets \cite{CTTW}
\be
R_\rho^{_{\rm PT}}(e) = \vev{\theta(e - \rho(N))}_{_{\rm PT}} =
\VEV{\theta\left(e - \frac{M_R^2}{Q^2}\right)}_{_{\rm PT}}
\VEV{\theta\left(e - \frac{M_L^2}{Q^2}\right)}_{_{\rm PT}}\,.
\label{R-rho}
\ee
The perturbative expressions \re{R-t} and \re{R-rho} are valid in the two-jet
kinematical region $\Lambda_{_{\rm QCD}}/Q \ll e < e_{\rm max}$ except the
end-point region $e\sim\Lambda_{_{\rm QCD}}/Q$, in which the energy of emitted
soft particles scales as $k_\perp\sim e Q\sim\Lambda_{_{\rm QCD}}$ and
perturbation theory is expected to fail.

Calculating the radiator functions $R_e(e)$ one has to combine together
perturbative and nonperturbative corrections. In the case of inclusive
distributions, like deep inelastic structure functions and Drell-Yan
distributions, this can be achieved by applying the factorisation theorems. They
allow to separate short-distance dynamics into perturbatively calculable
coefficient functions and absorb large-distance corrections into universal
nonperturbative distributions. Specific feature of the differential event-shape
distributions is that they are not inclusive quantities but rather weighted
cross-sections and, as a consequence, the standard methods are not applicable in
this case.

It turns out \cite{KS-shape} that IR factorization still holds for the leading
term $\sigma_0$ in the expansion of the event-distributions \re{sigma_0} in the
end-point region $e\sim\Lambda_{\rm QCD}/Q$. Its origin has a simple physical
interpretation. In end-point region, the final state in $\e^+\e^--$annihilation
consists of two narrow jets surrounding by a cloud of soft gluons.
Nonperturbative corrections $\sim 1/(Q^2e)$ and $\sim 1/(Q e)$ are associated
with emission of collinear particles with the transverse momenta $k_\perp^2 \sim
Q^2 t$ and soft particles on the energy scale $k_\perp\sim Q e$, respectively.
Neglecting power corrections to \re{sigma_0} on a larger scale, $\sim 1/(Q^2e)$,
we may restrict analysis to soft particles only. Since soft particle cannot
resolve the internal structure of narrow jets of transverse size $k_\perp^2
\sim Q^2 e$, we may effectively replace two jets by a pair
of energetic quark and antiquark moving back-to-back with the energy $\sim Q/2$.
The internal dynamics of two jets is governed by perturbative branching of quark
and antiquark while effects of their interaction with soft gluons can be
factorized out into the eikonal phase $W_+W_-^\dagger$ with $W_+$ and $W_-$ being
the eikonal phases of quark and antiquark, respectively. They are given by Wilson
lines $W_\pm=P\exp(i\int_0^\infty d s n_\pm A(sn_\pm))$ in which soft gluon field
$A_\mu(x)$ is integrated along the light-like directions $n_\pm$ defined by the
momenta of two outgoing jets. In the end-point region, collinear and soft
particles provide additive contributions to the shape variables
$e=(t\,,\rho\,,C)$. As a consequence, the radiator functions are given in all
three cases by a convolution of perturbative radiators $R_{_{\rm PT}}$ and the
same universal nonperturbative distribution $f(\varepsilon_R,\varepsilon_L)$
describing the energy flow into the right and left hemispheres in the final
state, $\varepsilon_R$ and $\varepsilon_L$, respectively, created by
nonperturbative soft gluon radiation. The nonperturbative distribution
$f(\varepsilon_R,\varepsilon_L)$ is defined as follows \cite{KS-shape}
\be
f(\varepsilon_R,\varepsilon_L)= \sum_{N} |\vev{0|W_+ W_-^\dagger|N}|^2 \delta(
\varepsilon_{_R}-(k_{_R}n_+))\delta( \varepsilon_{_L}-(k_{_L}n_-))\,.
\label{shape}
\ee
Here, sum goes over all possible soft gluon final states $\ket{N}$ with $k_R$ and
$k_L$ being the total momentum of soft particles moving into right and left
hemispheres, respectively. The quantities $(k_{_R}n_+)$ and $(k_{_L}n_-)$ define
the projection of the soft gluon momenta onto the directions of two jets,
$n_\pm^\mu=(1,{\mathbf 0}_\perp,\pm 1)$, propagating into the same hemisphere.

Finally, the factorized expressions for the radiator function for the $t-$ and
$C-$variables look like
\ba
R_t(e) &=& \int_0^{e Q} d\varepsilon \, f_t(\varepsilon)\, R_t^{_{\rm
PT}}\lr{e-\frac{\varepsilon}{Q}}
\label{R-t-f}
\\
R_C(e) &=& \int_0^{\frac2{3\pi}e Q} d\varepsilon \, f_t(\varepsilon)\,
R_C^{_{\rm PT}}\lr{e-\frac{3\pi}2\frac{\varepsilon}{Q}}
\label{R-C-f}
\ea
with nonperturbative distribution $f_t(\varepsilon)$ defined as
\be
f_t(\varepsilon) = \int d\varepsilon_R\int d\varepsilon_L
f(\varepsilon_R,\varepsilon_L) \,\delta(\varepsilon-\varepsilon_R-\varepsilon_L)
=\int_0^\varepsilon d \varepsilon'\, f(\varepsilon-\varepsilon',\varepsilon')\,.
\label{f-sing}
\ee
In the case of the $\rho-$variable,
\be
R_\rho(e) = \int_0^{e Q} d\varepsilon_R\int_0^{e Q} d\varepsilon_L  \,
f(\varepsilon_R,\varepsilon_L) R_J^{_{\rm PT}}\lr{e-\frac{\varepsilon_R}{Q}}
R_J^{_{\rm PT}}\lr{e-\frac{\varepsilon_L}{Q}}
\label{R-rho-f}
\ee
with $R_\rho^{_{\rm PT}}(e)=[R_J^{_{\rm PT}}(e)]^2$ and $R_J^{_{\rm PT}}(e)
=\VEV{\theta\left(e - {M_R^2}/{Q^2}\right)}_{_{\rm PT}}$ being a single jet
radiator function. We would like to stress that Eqs.~\re{R-t-f}, \re{R-C-f} and
\re{R-rho-f} hold in the region $\Lambda_{\rm QCD}^2/Q^2 < e < e_{\rm max}$. They resum all
power corrections of the form $1/(Q e)^n$ and are valid up to corrections $\sim
1/(Q^2 e)$. According to \re{R-t-f}, \re{R-C-f} and \re{R-rho-f}, the power
corrections have a different form for $\rho$ and $e=(t,C)$ variables. In the
latter case, the radiator function depends on an overall energy flowing into both
hemispheres and described by the integrated distribution \re{f-sing}.

Nonperturbative corrections to the radiator functions \re{R-t-f}, \re{R-C-f} and
\re{R-rho-f} are governed by the universal shape function $f(\varepsilon_R,\varepsilon_L)$.
This function is different from the well-known inclusive QCD distributions and
its operator definition was given in \cite{KS-shape}. Using \re{shape} it is
straightforward to show that $f(\varepsilon_R,\varepsilon_L)$ is a symmetric
function of its arguments, it does not depend on the center-of-mass energy $Q$
and is normalized as
\be
\frac{d}{d Q^2}f(\varepsilon_R,\varepsilon_L)=0\,,\qquad
\int d\varepsilon_R \int d\varepsilon_L\, f(\varepsilon_R,\varepsilon_L)=1\,,
\label{props}
\ee
where the last relation follows from unitarity of the eikonal phase $W_+
W_-^\dagger$. The matrix element entering \re{shape} does not depend on any
kinematical scale and, as a consequence, the momenta of soft gluons contributing
to \re{shape} are not restricted from above. To separate the region of small
gluon momenta one has to introduce the factorisation scale $\mu$. Then, the shape
function describes the contribution of gluons with $k_\perp < \mu$, while the
contribution of gluons with $k_\perp>\mu$, is absorbed into perturbative
radiator function $R(e)$. In this way, both nonperturbative shape function and
perturbative radiator become $\mu-$dependent while this dependence cancel in
their convolution \re{R-t-f},
\re{R-C-f} and \re{R-rho-f}. Since the $\mu-$dependence of radiator function
$R_{_{\rm PT}}$ can be calculated perturbatively, the above condition allows to
obtain the evolution equations on the nonperturbative distributions
\cite{KS-shape}. Clearly, there exists an ambiguity in implementing IR cut-off
inside perturbative expressions. Different prescriptions correspond to different
ways of regularizing IR renormalon singularities and therefore lead to the
different expressions for the nonperturbative distributions. In what follows we
shall impose a ``hard'' IR cut-off \cite{KS-shape} on gluon momenta inside the
perturbative radiator functions entering \re{R-t-f}, \re{R-C-f} and \re{R-rho-f}
as \cite{KS-shape}
\be
R_{_{\rm PT}}(e)\to R_{_{\rm PT}}(e;\mu) = \theta\lr{e-\frac{\mu}{Q}} R_{_{\rm
PT}}^{^{\rm NLL}}(e) +
\theta\lr{\frac{\mu}{Q} -e} R_{_{\rm PT}}^{^{\rm NLL}}(\mu/Q)\,.
\label{R-fin}
\ee
Throughout the paper we shall substitute $R_{_{\rm PT}}^{^{\rm NLL}}(e)$ by its
perturbative expression resummed to the NLL accuracy and matched into two-loop
explicit expressions within the modified $\ln R-$matching scheme
\cite{CTTW}. Thus defined radiator function \re{R-fin} depends on two scales,
$\Lambda_{\rm QCD}$ and IR cut-off $\mu$, that we choose as
\be
\Lambda_{\rm QCD} = \mu = 0.25 \ {\rm GeV}\,.
\label{mu}
\ee
Within the prescription \re{R-fin}, the ``regularized'' perturbative spectrum
$d\sigma_{_{\rm PT}}(e;\mu)\/d e= d R_{_{\rm PT}}(e;\mu)/d e$ coincides with the
$\ln R-$matched perturbative distribution $dR_{_{\rm PT}}^{^{\rm NLL}}(e)/de$ for
$\mu/Q < e<e_{\rm max}$ and it vanishes inside the nonperturbative ``window''
$0<e<\mu/Q$. Choosing the value of $\mu$ in \re{mu} one has to be sure that the
end-point of the perturbative distribution, $e=\mu/Q$, belongs to applicability
range of the NLL resummed radiator function $R_{_{\rm PT}}^{^{\rm NLL}}(e)$
\cite{CTTW}, $2\beta_0\alpha_s(Q^2)\ln e < 1$. Despite the fact that the
perturbative spectrum is well defined at $e=\mu/Q$ we do not expect that it
provides a reasonable description of the physical distribution in the end-point
region. Indeed, it is in this region that nonperturbative power corrections
become dominant.

\section{Differential distributions}

Differentiating the radiator functions \re{R-t-f} and \re{R-C-f} we obtain the
following expressions for the differential $t-$distribution
\be
\frac1{\sigma_{\rm tot}}\frac{d\sigma_t}{d e}= Qf(Qe;\mu) R_t^{_{\rm PT}}(0;\mu)
+\int_0^{Qe} d\varepsilon f_t(\varepsilon;\mu) \frac{d\sigma_t^{_{\rm PT}}
(e-\varepsilon/Q;\mu)}{d e}
\label{dis-t}
\ee
and $C-$distribution
\be
\frac1{\sigma_{\rm tot}}\frac{d\sigma_C}{d e}= \frac2{3\pi}Qf\lr{\frac{3\pi}2Q e;\mu}
R_C^{_{\rm PT}}(0;\mu)+\int_0^{\frac2{3\pi} Qe} d\varepsilon f_t(\varepsilon;\mu)
\frac{d\sigma_C^{_{\rm PT}} \lr{e-\frac{3\pi}2\frac{\varepsilon}{Q};\mu}}{d e}\,.
\label{dis-C}
\ee
Here, we indicated explicitly the dependence of nonperturbative shape function
and perturbative distributions on the factorization scale $\mu$. Two terms
entering the r.h.s.\ of \re{dis-t} and \re{dis-C} have the following
interpretation. Since the shape function $f_t(\varepsilon)$ rapidly vanishes for
large $\varepsilon$, the first term contributes inside the nonperturbative
window $0 \le e < \mu/Q$. In this region the emission of perturbative real soft
gluons is suppressed due to cut-off imposed on soft gluon momenta $k_\perp > \mu$
and the shape of the distribution is governed entirely by nonperturbative
function $f_t(\varepsilon)$. Additional Sudakov factor $R^{_{\rm PT}}(0;\mu)$
takes into account the contribution of virtual soft gluons with $\mu < k_\perp
<Q$ and it rapidly vanishes as $\mu$ decreases. The second term in \re{dis-t} and
\re{dis-C} defines the spectrum inside the perturbative window $\mu/Q < e< e_{\rm
max}$. In this region, nonperturbative corrections smear the perturbative
spectrum over the interval $\Delta e \sim \Lambda_{_{\rm QCD}}/Q$.

For the heavy mass distribution one finds
\be
\frac1{\sigma_{_{\rm tot}}}\frac{d\sigma_\rho}{d e}
=Q f_\rho(e Q,e Q;\mu)R_J^{_{\rm PT}}(0;\mu) +
\int_0^{e Q} d\varepsilon\,f_\rho(\varepsilon,e Q;\mu)
\frac{d\sigma_J^{_{\rm PT}}
(e-\varepsilon/Q;\mu)}{d e}\,,
\label{dis-rho}
\ee
where $d\sigma_J^{_{\rm PT}}/d e$ is single jet distribution resummed to the NLL
order and defined by the radiator function \re{R-rho-f}, $d\sigma_J^{_{\rm PT}}/d
e= d R_J^{_{\rm PT}}(e)/de$. The heavy mass nonperturbative distribution is given
by
\be
f_\rho(\varepsilon,e Q) = 2 \int_0^{e Q} d\varepsilon'\,
f(\varepsilon,\varepsilon') R_J^{_{\rm PT}}\lr{e-\frac{\varepsilon'}{Q}}\,.
\label{fH}
\ee
Comparing \re{dis-rho} with \re{dis-t} and \re{dis-C} we notice that the
factorized expressions for the differential $t-$, $C-$ and $\rho-$distributions
have a similar form but the structure of power corrections is different in the
case of the heavy mass. In distinction with \re{f-sing}, the heavy mass
nonperturbative function $f_\rho$ depends on the shape variable, $e$, and the
center-of-mass energy, $Q$. This dependence is controlled by perturbative
radiator function and has the following interpretation. In the two-jet limit,
the invariant mass of each jet is given by the sum of perturbative and
nonperturbative contributions, $M_R^2=M^2_{R,{\rm PT}} +\varepsilon_RQ$ and
$M_L^2=M^2_{L,{\rm PT}} +\varepsilon_L Q$. Perturbative radiation leads to
$M^2_{R,{\rm PT}}/Q^2\sim M^2_{L,{\rm PT}}/Q^2\sim {\cal O}(\alpha_s(Q))$, while
nonperturbative contribution scales as $\varepsilon_R\sim\varepsilon_L\sim{\cal
O}(\Lambda_{\rm QCD})$. In contrast with the $t-$variable, which depends on the
sum of both masses and therefore is additive with respect to perturbative and
nonperturbative contributions, the $\rho-$parameter is defined by the largest
mass for which the ``additivity'' property is lost. Namely, comparing invariant
masses flowing into two hemispheres one encounters a situation when masses of
two perturbative jets are of the same order, $M^2_{R/L,_{\rm PT}}={\cal O}(Q^2)$,
while their difference is much smaller $|M_{L,{\rm PT}}^2-M_{R,{\rm PT}
}^2|={\cal O}(Q\Lambda_{\rm QCD})$.%
\footnote{To see that this configuration is not rare it is enough to notice that
it corresponds to the vicinity of peak of the perturbative distribution over the
difference of the jet masses $|M_L^2-M_R^2|/Q^2$ \cite{KNMW}.} In this case,
nonperturbative correction to the difference of the jet masses becomes
comparable with the perturbative contribution $|M_{L,{\rm PT}}^2-M_{R,{\rm
PT}}^2|\sim Q |\varepsilon_L-\varepsilon_R|$, and therefore it can invert the
perturbative hierarchy of jet masses, $M_{R,{\rm PT}}^2 < M_{L,{\rm PT}}^2$,
into $M_R^2 > M_L^2$, for instance. Expression \re{dis-rho} takes into account
this effect through the induced $Q-$dependence of the nonperturbative function
\re{fH}.

According to \re{dis-t}, \re{dis-C} and \re{dis-rho} the nonperturbative
corrections to the $t-$, $C-$ and $\rho-$distributions involve two different
nonperturbative functions. They are related however to the same universal
nonperturbative shape function \re{shape} describing the energy flow into two
hemispheres in the final state. We recall, that
$f(\varepsilon_R,\varepsilon_L;\mu)$ depends on the cut-off $\mu$ imposed on the
maximal momenta of soft particles but it is independent on the center-of-mass
energy $Q$. By the definition, $f(\varepsilon_R,\varepsilon_L;\mu)$
distinguishes between particles propagating into right and left hemispheres in
the final state and therefore it is not completely inclusive with respect to
partonic final states. Namely, it takes into account that quarks and gluons
produced at short distances $\sim 1/Q$ and moving into one of the hemispheres
will eventually decay at large distances $\sim 1/\Lambda_{\rm QCD}$ and their
remnants could flow into opposite hemispheres.\footnote{Similar effect has been
studied using the IR renormalon approach in \cite{NS}.} This implies that,
firstly, in contrast with the well-known inclusive QCD distributions, the shape
function $f(\varepsilon_R,\varepsilon_L;\mu)$ is not related to the short
distance QCD dynamics and, in particular, its moments can not be related to
hadronic matrix elements of {\it local\/} composite operators. Indeed, according
to the operator definition proposed in \cite{KS-shape}, the shape functions are
defined in terms of the so-called ``maximally nonlocal'' QCD operators
\cite{Nonloc,CS}. Secondly, non-inclusive corrections to the shape function
describe a ``cross-talk'' between two hemispheres in the final state leading to
correlations between $\varepsilon_R$ and $\varepsilon_L$. As a consequence, the
shape function is not factorizable into the product of functions depending on
the energy flowing into separate hemispheres
\be
f(\varepsilon_R,\varepsilon_L)=f_{\rm incl}(\varepsilon_R)f_{\rm
incl}(\varepsilon_L)+\delta f_{\rm non-incl} (\varepsilon_R,\varepsilon_L)\,.
\label{non-inc}
\ee
One should notice that similar property holds for perturbative Sudakov resummed
radiator function \re{R-rho}. However, one finds that there the factorization
holds to the NLL accuracy, Eq.~\re{R-rho}, and non-inclusive corrections first
appear at the NNLL level $\sim\alpha_s^2 (\alpha_s L)^N$.

In what follows we shall rely on a particular ansatz for the shape function
$f(\varepsilon_R,\varepsilon_L)$ which agrees with general properties of
nonperturbative QCD distributions and has been used in previous studies of power
corrections to the thrust distributions \cite{KS-shape}. Namely, one expects that
for small values of $\varepsilon_R$ and $\varepsilon_L$ the shape function
should vanish as a power of the energy. Similarly,
$f(\varepsilon_R,\varepsilon_L)$ should rapidly vanish as $\varepsilon_R$ or
$\varepsilon_L$ becomes large. Taking into account these properties together with
\re{non-inc} one chooses the following expression
\be
f(\varepsilon_R,\varepsilon_L)=\frac{{\cal N}(a,b)}{\Lambda^2}
\lr{\frac{\varepsilon_R\varepsilon_L}{\Lambda^2}}^{a-1}
\exp\lr{-\frac{\varepsilon_R^2+\varepsilon_L^2+2b \varepsilon_R\varepsilon_L}{\Lambda^2}}
\,.
\label{ansatz}
\ee
It depends on two dimensionless parameters $a$ and $b$ and the scale $\Lambda$.
The factor ${\cal N}(a,b)$ is fixed by normalization condition
\re{props}.

The free parameters, $a$, $b$ and $\Lambda$, have the following meaning. The
exponent $a$ determines how fast the shape function vanishes at the origin. The
scale $\Lambda$ sets up the typical energy of soft radiation. The parameter $b$
controls the non-inclusive contribution to the shape function and its possible
values are restricted as $b>-1$ in order for the shape function
\re{ansatz} to be normalizable. Non-inclusive corrections vanish at $b=0$,
$\delta f_{\rm non-incl} =0$ in \re{non-inc}. For $b\gg 1$, the shape function
enhances the regions of the phase space $\varepsilon_R\gg\varepsilon_L$ and
$\varepsilon_R\ll\varepsilon_L$, in which most of the energy flows into one of
the hemispheres. For $b\to -1$ the energies are of the same order,
$\varepsilon_R\sim\varepsilon_L$, and strongly correlated to each other. We
expect that non-inclusive corrections to the shape function should be important
and the configurations in which energy flows mostly into one of the hemispheres
to be suppressed. This suggests that the possible values of the $b-$parameter
should lie within the interval $-1 < b < 0$.

\begin{figure}[ht]
\begin{center}
\hspace*{-5mm}
\epsfig{file=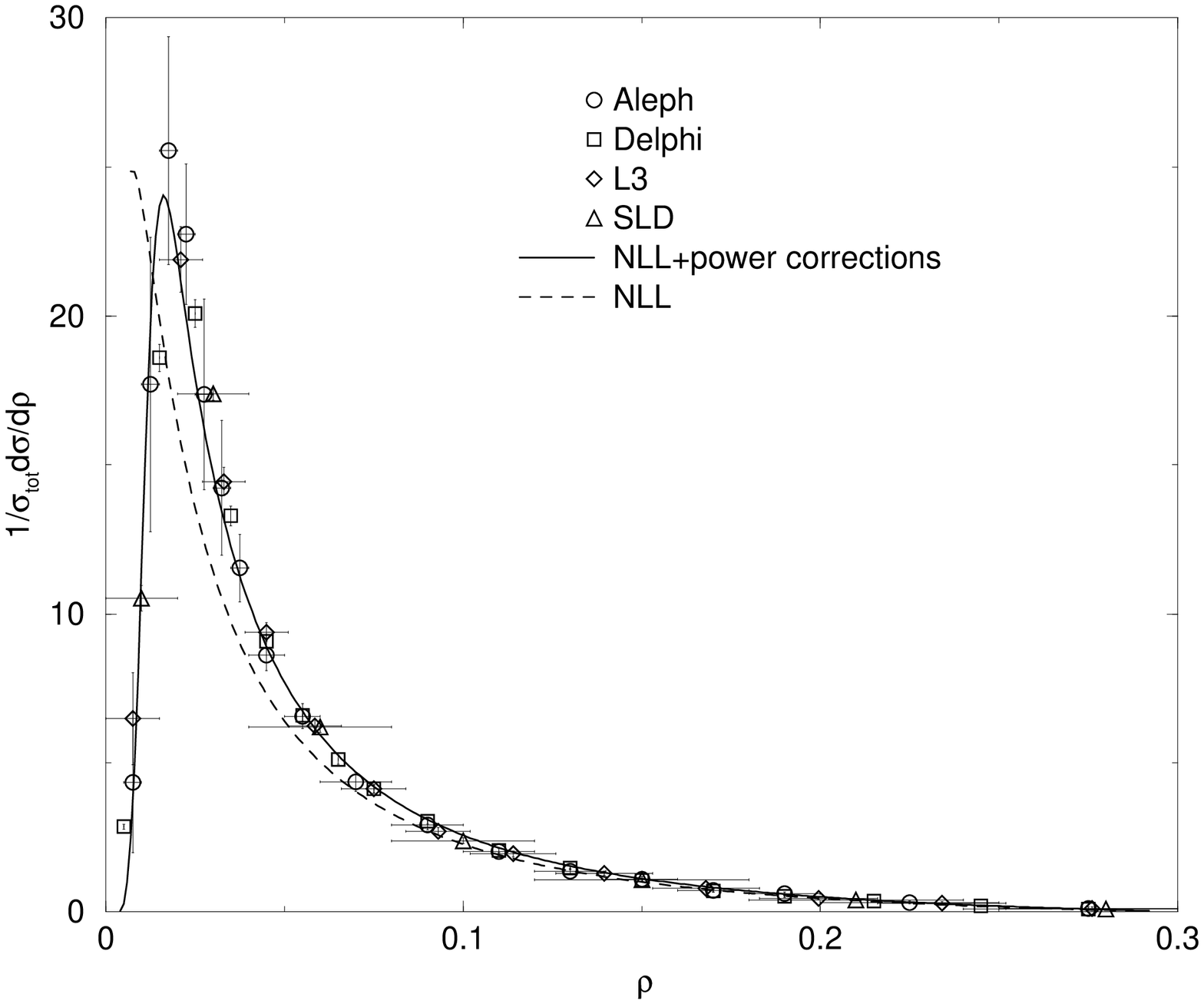,angle=-90,width=7.5cm}
\hspace*{5mm}
\epsfig{file=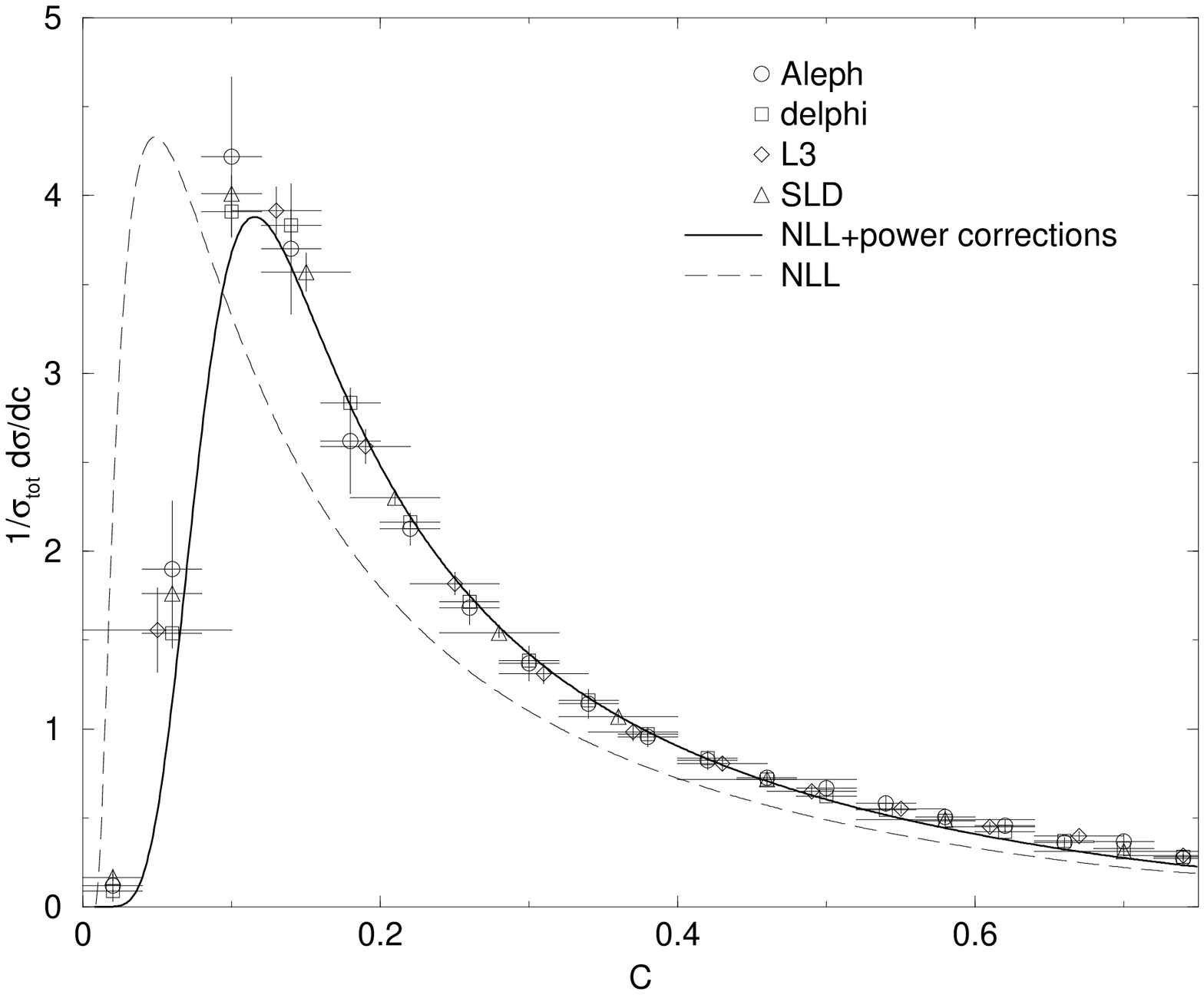,angle=-90,width=7.5cm}
\\
{\small (a)}\hspace*{75mm}{\small (b)}
\end{center}
\caption[]{Heavy jet mass (a) and $C-$parameter (b) distributions at $Q=M_Z$ with and without
power corrections included.}
\label{Fig-91}
\end{figure}

The parameters $a$, $b$ and $\Lambda$ depend on the factorization scale $\mu$
and are independent on the center-of-mass energy $Q$ as well as the choice of
the shape variable $e=(t,\rho,C)$. This allows to fit their values by comparing
the event shape distributions, Eqs.~\re{dis-t}, \re{dis-C} and \re{dis-rho}, with
the most precise available experimental data at $Q=M_Z$. Following this procedure
we found that the fit to the heavy jet mass distribution is more sensitive to
the choice of the parameters (especially to the non-inclusiveness parameter $b$)
then the one to the thrust and the $C-$parameter. Then, fitting the heavy jet
mass distribution at $Q=M_Z$ as shown in Fig.~\ref{Fig-91}a we obtain
\be
a=2\,,\qquad b=-0.4\,,\qquad \Lambda=0.55 \ {\rm GeV}\,.
\label{para}
\ee
Using these values we compare the QCD predictions for the $C-$parameter
distribution at $Q=M_Z$ with and without nonperturbative corrections included as
shown in Fig.~\ref{Fig-91}b. Similar plot for the thrust distribution can be
found in \cite{KS-shape}. We observe that the differential distributions
\re{dis-rho} and \re{dis-C} combined with the shape function, Eqs.~\re{ansatz}
and \re{para}, correctly describe the data throughout the interval $0<e<e_{\rm
max}$ including the end-point region $e={\cal O}(\Lambda_{\rm QCD}/Q)$. In
addition, the $\rho-$parameter distribution turns out to be very sensitive to
the choice of the $b-$parameter. The fact that its value, \re{para}, is
relatively large indicates that non-inclusive corrections to the shape function
\re{non-inc} are important indeed.

\begin{figure}[ht]
\begin{center}
\hspace*{-10mm}
\epsfig{file=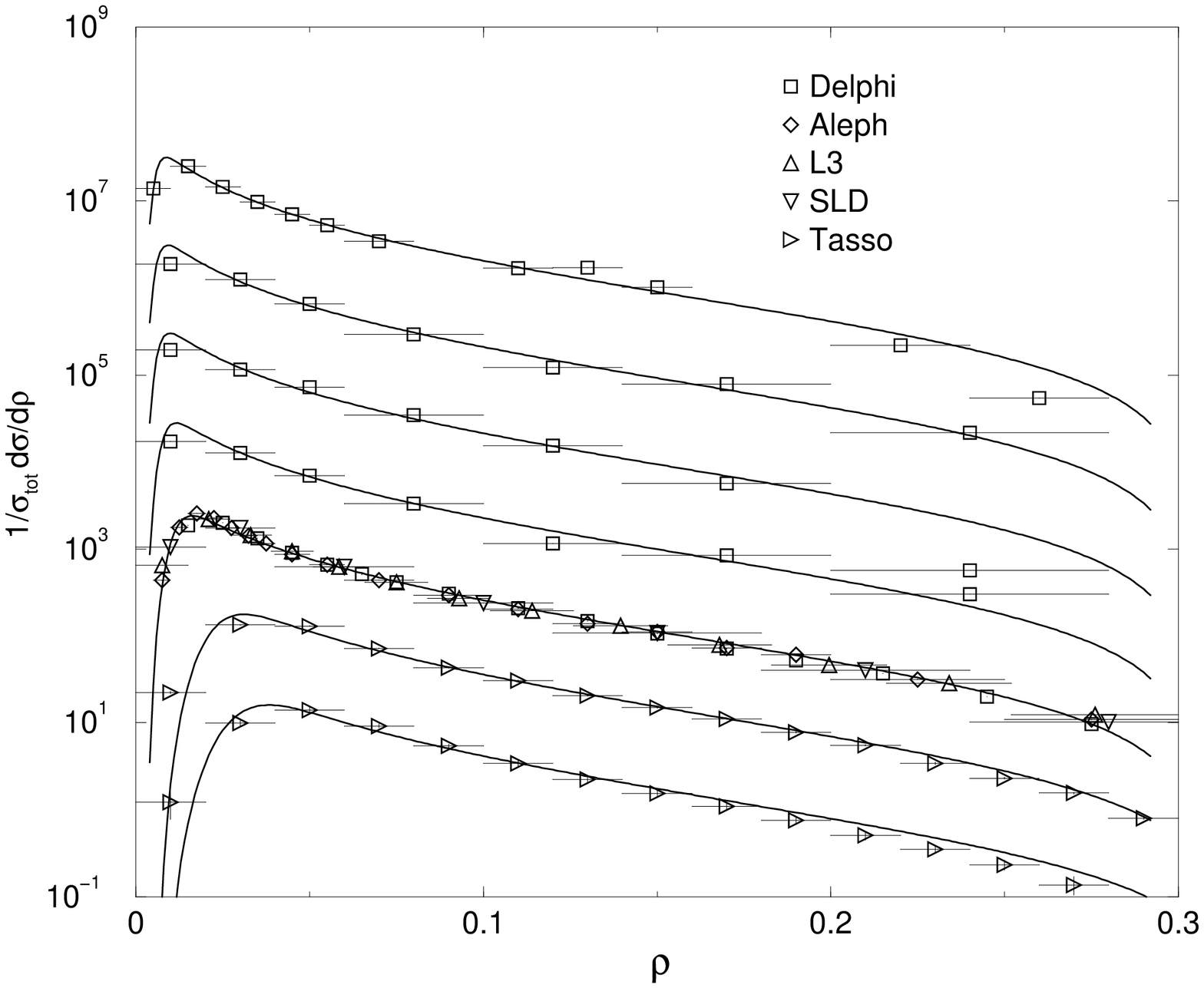,angle=-90,width=8cm}
\epsfig{file=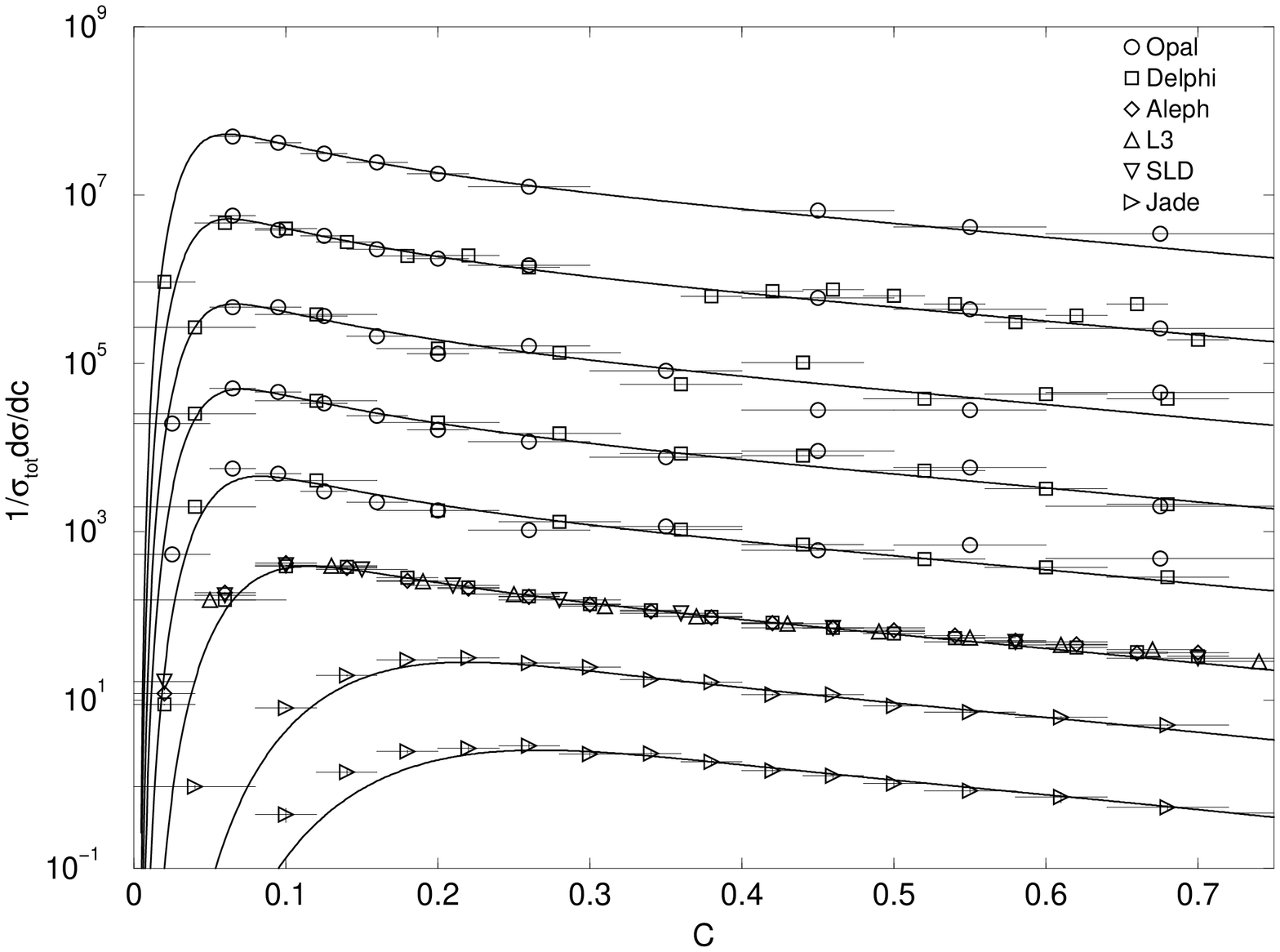,angle=-90,width=88mm}
\\
{\small (a)}\hspace*{80mm}{\small (b)}
\end{center}
\caption[]{Comparison of the QCD predictions for the heavy jet mass (a) and
$C-$parameter (b) distributions with the data at different center-of-mass
energies (from bottom to top): $Q/{\rm GeV} = 35\,, 44\,, 91\,, 133\,, 161\,,
172\,, 183\,,189$, based on the shape function.}
\label{Fig-C}
\end{figure}

Having determined the parameters of the shape function, Eq.~\re{para}, at the
reference energy scale $Q=M_Z$, we can now apply the factorized expressions for
the differential distributions, \re{dis-t}, \re{dis-C} and \re{dis-rho} with the
{\it same\/} ansatz for the shape function \re{ansatz} to obtain the QCD
predictions at different energy and compare them with the data. The combined
plot for the $\rho-$ and $C-$parameter distributions over the center-of-mass
energy interval $35\ {\rm GeV}\le Q \le 189\ {\rm GeV}$ is shown in
Fig.~\ref{Fig-C} a and b, respectively. Similar plot for the thrust distribution
can be found in
\cite{KS-shape}. We observe that the theoretical curves reproduce the data over the whole
interval of the shape variables including the end-point region.

\section{Moments of the event shapes}

Recently, the experimental data for the first few moments of various event shape
distributions became available \cite{Exp}. Their analysis indicates a presence
of large hadronization corrections whose form deviates from IR renormalon models
describing the nonperturbative corrections to the distributions as the shift of
perturbative spectrum.

Let us apply the obtained expressions for differential distributions to calculate
the first two moments of the $t-$, $C-$ and $\rho-$distributions defined as
\be
\vev{e^n} = \int_0^{e_{\rm max}} de\, e^n \frac1{\sigma_{\rm tot}}\frac{d\sigma}{d
e}\,, \qquad (n=1\,,\ 2)\,.
\label{moms}
\ee
Here, integration goes only over the part of the available phase space, $0<e<
e_{\rm max}$, corresponding to the three-particle final states, and it does not
take into account the contribution of multi-jet final states, $e>e_{\rm max}$.
Quantitative description of hadronization corrections to such final states is not
available yet. Putting an upper limit on the value of the shape variable in
\re{moms} allows us to avoid the latter contribution and to replace the differential
distribution ${d\sigma}/{d e}$ in \re{moms} by the obtained expressions
\re{dis-t}, \re{dis-C} and \re{dis-rho} which are valid for $0< e< e_{\rm max}$.

Using general expression \re{factor} one calculates the mean value of the event
shape as
\be
\vev{e} = \vev{e}_{_{\rm PT}} + \frac{\vev{\varepsilon}}{Q}\left[1-{e_{\rm
max}}\frac{d\sigma_{_{\rm PT}}(e_{\rm max})}{de}\right] +{\cal
O}\lr{\frac1{Q^2}}\,,
\label{mean-gen}
\ee
where $\vev{...}_{_{\rm PT}}=\int_0^{e_{\rm max}} de\,(...)\, {d\sigma_{_{\rm
PT}}}/{d e}$ denotes averaging with respect to perturbative distribution and the
scale $\vev{\varepsilon}$ is defined as the first moment of the shape function,
$\vev{\varepsilon}=\int d\varepsilon\,\varepsilon f(\varepsilon)$. It is
important to remember that the factorized expressions for the differential
distributions \re{dis-t}, \re{dis-C} and \re{dis-rho} are valid up to ${\cal
O}(1/(Q^2 e))-$correc\-tions which may modify the mean value $\vev{e}$ by ${\cal
O}(1/Q^2)-$terms. The additional factor in front of $\vev{\varepsilon}/Q$ takes
into account that close to the edge of the three-particle phase space, $ e\to
e_{\rm max}$, the perturbative distribution \re{A_e} could take nonzero values
\be
{e_{\rm max}}\frac{d\sigma_{_{\rm PT}}(e_{\rm max})}{de}=
\frac{\alpha_s(Q)}{2\pi}
e_{\rm max}A_e(e_{\rm max}) + {\cal O}(\alpha_s^2)\,.
\ee
This can be checked using the explicit expressions for perturbative
distributions \re{A's}. We find that ${d\sigma_{_{\rm PT}}}/{de}$ vanishes to
one-loop order as $e\to e_{\rm max}$ for the $t-$ and $\rho-$variables while for
the $C-$parameter it approaches a finite value. Finally, calculating the mean
values $\vev{\varepsilon}$ with respect to nonperturbative distributions
$f_t(\varepsilon)$ and $f_\rho(\varepsilon,eQ)$ defined in \re{f-sing} and
\re{fH}, respectively, we obtain
\ba
\vev{t} &=& \vev{t}_{_{\rm PT}} + \frac{\lambda_1}{Q} + {\cal O}(1/Q^2)\,,
\label{mean-t}
\\
\vev{\rho} &=& \vev{\rho}_{_{\rm PT}} + \frac{\lambda_1}{2Q} + {\cal O}(1/Q^2)\,.
\label{mean-rho}
\ea
Similarly, for the mean value of the $C-$parameter we get
\be
\vev{C} = \vev{C}_{_{\rm PT}} + \frac{3\pi}{2} \frac{\lambda_1}{Q}
\left[1-\frac{\alpha_s(Q)}{2\pi}5.73 + {\cal O}(\alpha_s^2) \right]+ {\cal
O}(1/Q^2)\,.
\label{mean-C}
\ee
\begin{figure}[ht]
\begin{center}
\epsfig{file=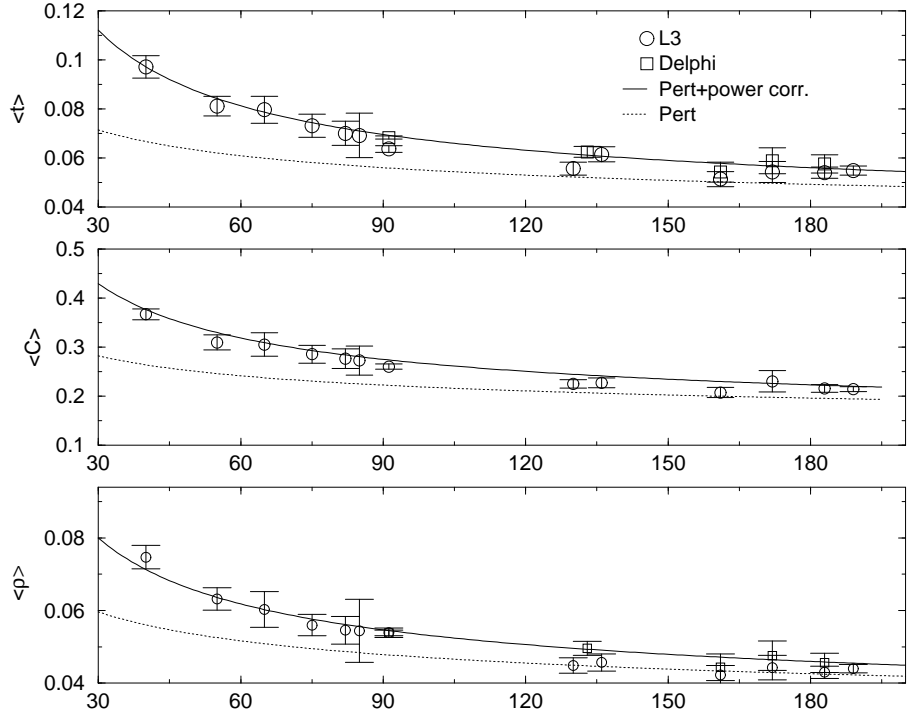,angle=-90,width=12cm}
\end{center}
\caption[]{Comparison of the QCD predictions to the mean values $\vev{t}$,
$\vev{\rho}$ and $\vev{C}$ with the data. Dotted lines denote ${\cal
O}(\alpha_s^2)-$perturbative contribution, solid lines take into account power
corrections given by Eqs.~\re{mean-t}, \re{mean-rho} and \re{mean-C}.}
\label{Fig-mean}
\end{figure}
Here, large perturbative coefficient originates from \re{A's} and it reduces a
magnitude of the nonperturbative scale $\lambda_1$ by $11\%$ at $Q=M_Z$.
Relations \re{mean-t} and \re{mean-rho} coincide with IR renormalon model
predictions \cite{W,AZ,CW}, while \re{mean-C} differs by perturbative
$\alpha_s(Q)-$dependent ``boundary" term. Nonperturbative $Q-$independent scale
$\lambda_1$ is given by
\be
\lambda_1 = \int d\varepsilon_R \int d\varepsilon_L\, (\varepsilon_R+\varepsilon_L)
f(\varepsilon_R,\varepsilon_L) = \int d\varepsilon\, \varepsilon f_t(\varepsilon)
\,.
\label{lambda_1}
\ee
Substituting expression for the shape function, Eq.~\re{ansatz}, one finds
$\lambda_1=\Lambda\,\varphi(a,b)$ with $\varphi$ given by
${}_2F_1-$hypergeometric series. Using the values of the parameters \re{para} we
find
\be
\lambda_1 = 1.22 \ {\rm GeV}\,.
\ee
We would like to recall that this value depends on the factorization scale
$\mu$, Eq.~\re{mu}, and its $\mu-$dependence is described by QCD evolution
equation \cite{KS-shape}. Obviously, the value of $\lambda_1$, and as a
consequence $1/Q-$corrections to the mean values \re{mean-t}, \re{mean-rho} and
\re{mean-C} are less sensitive to the choice of the parameters $a$, $b$ and
$\Lambda$ as compared with nonperturbative corrections to the corresponding
differential distributions.

The comparison of the QCD predictions, \re{mean-t}, \re{mean-rho} and
\re{mean-C}, with the data over the energy interval $35\ {\rm GeV} \le Q \le 189\
{\rm GeV}$ is shown in Fig.~\ref{Fig-mean}. One should notice that \re{mean-t},
\re{mean-rho} and \re{mean-C} describe the contribution of the two-jet configurations
while experimental data take into account all possible final states. A good
agreement observed in Fig.~\ref{Fig-mean} indicates that the contribution to the
mean values of the final states with three and more jets as well as ${\cal
O}(1/(Q^2e))$ subleading corrections to the distributions are subdominant.
Indeed, the dominant contribution to the moments \re{moms} comes from the
vicinity of peak of the differential distribution $e^n d\sigma/de$. Using
existing experimental data one can show \cite{G-M} that for $n=1$ and $n=2$ the
position of the peak is located in the two-jet kinematical region while for
higher $n$ it moves towards larger $e$ for which the integral \re{moms} becomes
very sensitive to the choice of the upper integration limit $e_{\rm max}$. We
shall use this observation calculating the second moment of the event shape
distributions.

\begin{figure}[ht]
\begin{center}
\epsfig{file=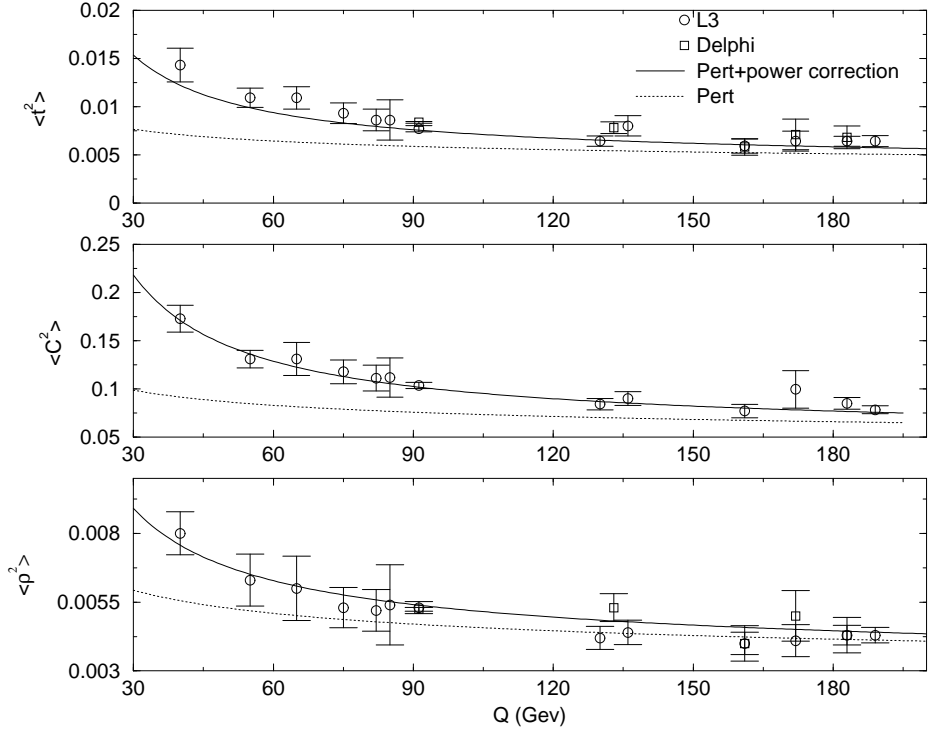,angle=-90,width=12cm}
\end{center}
\caption[]{Comparison of the QCD predictions for the second moments
$\vev{t^2}$, $\vev{\rho^2}$ and $\vev{C^2}$ with the data. Dotted lines denote
${\cal O}(\alpha_s^2)-$perturbative contribution, solid lines take into account
power corrections given by Eq.~\re{2nd}.}
\label{Fig-2nd}
\end{figure}

Let us apply \re{moms} to calculate $\vev{e^2}$. Neglecting $1/(Q^2
e)-$corrections to the distribution \re{factor} we find after some algebra the
following general expression
\ba
\vev{e^2} &=& \vev{e^2}_{_{\rm PT}}+\frac{\vev{\varepsilon}}{Q}\left[
2\vev{e}_{_{\rm PT}}-{e_{\rm max}^2}\frac{d\sigma_{_{\rm PT}}(e_{\rm
max})}{de}\right]+
\nonumber
\\
&+&\frac{\vev{\varepsilon^2}}{Q^2}\left[1-{e_{\rm max}}\frac{d\sigma_{_{\rm
PT}}(e_{\rm max})}{de} + \frac12{e_{\rm max}^2}\frac{d^2\sigma_{_{\rm PT}}(e_{\rm
max})}{de^2}
\right]+{\cal O}(1/Q^3)\,.
\ea
Similar to the mean value, \re{mean-gen}, the boundary terms vanish for the $t-$
and $\rho-$variables while for the $C-$parameter they provide a sizeable
contribution. Using the explicit expression for the shape functions,
\re{f-sing} and \re{fH}, we calculate the scales $\vev{\varepsilon^2}$
and take into account the boundary terms \re{A's} to obtain%
\footnote{We are grateful to O.~Biebel and S.~Kluth for providing us ${\cal
O}(\alpha_s^2)-$expressions for the moments of the event shapes.}
\ba
\vev{t^2}&=& \vev{t^2}_{_{\rm PT}} + 2\frac{\lambda_1}{Q}\vev{t}_{_{\rm
PT}}+\frac{\lambda_2}{Q^2}
\nonumber
\\
\vev{\rho^2}&=&\vev{\rho^2}_{_{\rm PT}} + \frac{\lambda_1}{Q}\vev{\rho}_{_{\rm
PT}}+\frac{\lambda_2+\delta\lambda_2(Q)}{4Q^2}
\label{2nd}
\\
\vev{C^2}&=& \vev{C^2}_{_{\rm PT}} + \frac{3\pi}2\frac{\lambda_1}{Q}
\left[2\vev{C}_{_{\rm PT}}- \frac{\alpha_s(Q)}{2\pi} 4.30 \right] +
\frac{9\pi^2}4\frac{\lambda_2}{Q^2}\left[1 - \frac{\alpha_s(Q)}{2\pi} 11.46\right]\,.
\nonumber
\ea
Here, the scale $\lambda_1$ was defined in \re{lambda_1} and new scales
$\lambda_2$ and $\delta\lambda_2$ are given by
\be
\lambda_2 = \vev{(\varepsilon_R+\varepsilon_L)^2}\,,
\qquad
\delta\lambda_2(Q) =
\langle \left(\varepsilon_R-\varepsilon_L\right)^2\rangle\,
\left\{1+4\int_0^{\rho_{\rm max}} d\rho'\rho'\,
\left({d\sigma^{\rm PT}_J\over d\rho'}\right)^2 \right\}\,,
\ee
where average is taken with respect to the shape function
$f(\varepsilon_R,\varepsilon_L)$. The $Q-$dependence of the scale
$\delta\lambda_2$ is attributed to perturbative prefactor depending on the single
jet distribution, $d\sigma_J^{\rm PT}/d\rho$, defined in \re{dis-rho}. Its origin
was explained in Sect.~4. We find that the value of this factor varies from
$2.19$ at $Q=10\, {\rm Gev}$ to $1.85$ at $Q=100\, {\rm Gev}$. It is important to
notice that $\vev{\left(\varepsilon_R-\varepsilon_L\right)^2}$ vanishes if one
does not take into account non-inclusive corrections to the shape function
\re{non-inc}, $\delta f_{\rm non-incl}=0$. Using \re{ansatz} and \re{para} one
gets
\be
\lambda_2 = 1.70 \ {\rm GeV}^2\,, \qquad
\vev{\left(\varepsilon_R-\varepsilon_L\right)^2}= 0.14 \ {\rm GeV}^2\,.
\ee
It follows from \re{2nd} that the boundary terms generate a sizable perturbative
corrections to the second moment of the $C-$parameter distribution and diminish
the magnitude of scales parameterizing $1/Q-$power corrections. Moreover, one
finds from \re{mean-t}, \re{mean-rho}, \re{mean-C} and \re{2nd} that the
variance of the distribution, $\vev{e^2}-\vev{e}^2$, does not receive $1/Q-$power
corrections for the $t-$ and $\rho-$variables while for the $C-$parameter the
boundary terms produce a negative $1/Q-$correction
\be
\vev{C^2}-\vev{C}^2=\vev{C^2}_{_{\rm PT}}-\vev{C}^2_{_{\rm PT}}
- 3.23\,\frac{\lambda_1}{Q} \alpha_s(Q)+ {\cal O}(1/Q^2)\,.
\ee
The comparison of the QCD predictions \re{2nd} with the experimental data is
shown in Fig.~\ref{Fig-2nd}. We would like to recall that the obtained
expressions for the moments do not take into account the contribution of
multi-jet final states configurations and assume a smallness of
$1/(Q^2e)-$correc\-tions to the distributions \re{sigma_0}. It is interesting to
note that the last assumption is supported by the recent analysis of the power
corrections to the first two moments of the thrust distribution in the single
dressed gluon approximation
\cite{G}. This analysis is complimentary to our studies since it does not resum
leading power corrections in the two-jet region and takes into account the
contribution coming from the region $e\to e_{\rm max}$.

\section{Conclusions}

In this paper we have studied the power corrections to the thrust, $t$, heavy
jet mass, $\rho$, and $C-$parameter distributions in the two-jet kinematical
region. Our analysis was based on the observation \cite{KS-shape} that
perturbative and nonperturbative effects can be separated in the differential
event shape distributions into calculable Sudakov resummed distribution,
$d\sigma_{\rm PT}/de$, and nonperturbative shape function,
$f(\varepsilon_R,\varepsilon_L)$, respectively. Each of them depends separately
on the factorization scale $\mu$ but this dependence cancels in their product.
The shape function describes the energy flow into two hemispheres in the final
state. It does not depend on the center-of-mass energy $Q$ as well as on the
choice of the event shape variable $e=t\,,\rho$ and $C$.

We demonstrated that away from the end-point region, $e\gg \Lambda_{\rm QCD}/Q$,
nonperturbative corrections to the distributions have a simple form \re{off-peak}
with the leading $1/Q-$power correction parameterized by a single scale given by
the first moment of the shape function. In this region, to which all performed
experimental analysis have been restricted so far, our predictions for the thrust
and heavy mass distributions and their mean values coincide with those of IR
renormalon based models while for the $C-$parameters we find an additional
sizeable perturbative contribution modifying the magnitude of the $1/Q-$power
correction \re{mean-C}.

In the end-point region, $e\sim \Lambda_{\rm QCD}/Q$, the obtained factorized
expressions for the distributions take into account power corrections of the form
$1/(Qe)^n$ for arbitrary $n$. They are controlled by the shape function through
\re{lambda_n} and are sensitive to the choice of this function. Comparing the QCD
predictions with the data we have chosen the simplest ansatz for the shape
function \re{ansatz} which is consistent with general properties of
nonperturbative distributions and includes nonzero correlations between energy
flows into different hemispheres. Examining the dependence of the distributions
on the corresponding parameter of the shape function we have observed that these
correlations play an important r\^ole and are not negligible.

\subsection*{Acknowledgments}

We would like to thank E.~Gardi and G.~Sterman for very interesting discussions.
We are grateful to O.~Biebel, G.~Salam and B.~Webber for useful correspondence.
This work was supported in part by the EU network ``Training and Mobility of
Researchers'', contract FMRX--CT98--0194 (G.K.) and the BFA fellowship -- Bourse
de coop\'eration Franco-Alg\'erienne (S.T.).


\begin{thebibliography}{99}

\bibitem{Exp}
O.~Biebel,
hep-ex/0006020; \\
M.~Acciarri {\it et al.}  [L3 Collaboration],
hep-ex/0005045; \\
G.~Abbiendi {\it et al.}  [JADE collaboration],
hep-ex/0001055; \\
G.~Dissertori,
Nucl.\ Phys.\ Proc.\ Suppl.\  {\bf 79} (1999) 438 [hep-ex/9904033]; \\
P.~Abreu {\it et al.}  [DELPHI Collaboration],
Phys.\ Lett.\  {\bf B456} (1999) 322.

\bibitem{Old}
F.~Barreiro,
Fortsch.\ Phys.\  {\bf 34} (1986) 503.

\bibitem{W}
B.~R.~Webber,
Phys.\ Lett.\  {\bf B339} (1994) 148 [hep-ph/9408222];
\\
Y.~L.~Dokshitzer and B.~R.~Webber,
Phys.\ Lett.\  {\bf B352} (1995) 451 [hep-ph/9504219].

\bibitem{KS}
G.~P.~Korchemsky and G.~Sterman,
Nucl.\ Phys.\  {\bf B437} (1995) 415 [hep-ph/9411211];
in Proceedings of the 30th Rencontres de Moriond: QCD and High Energy Hadronic
Interactions, France, 19-25 Mar 1995, Hadronic:0383-392 (QCD161:R4:1995:V.2)
[hep-ph/9505391].

\bibitem{AZ}
R.~Akhoury and V.~I.~Zakharov,
Nucl.\ Phys.\  {\bf B465} (1996) 295 [hep-ph/9507253].

\bibitem{BB}
M.~Beneke and V.~M.~Braun,
Nucl.\ Phys.\  {\bf B454} (1995) 253 [hep-ph/9506452].

\bibitem{Beneke}
M.~Beneke,
Phys.\ Rept.\  {\bf 317} (1999) 1 [hep-ph/9807443].

\bibitem{GG}
E.~Gardi and G.~Grunberg,
JHEP {\bf 9911} (1999) 016 [hep-ph/9908458].

\bibitem{DMW}
Y.~L.~Dokshitzer, G.~Marchesini and B.~R.~Webber,
Nucl.\ Phys.\  {\bf B469} (1996) 93 [hep-ph/9512336].

\bibitem{KS-shape}
G.~P.~Korchemsky,
in Proceedings of the 33rd Rencontres de Moriond, Les Arcs, France, 21-28 Mar
1998, pp.~489--498 [hep-ph/9806537];
\\
G.~P.~Korchemsky and G.~Sterman,
Nucl.\ Phys.\  {\bf B555} (1999) 335 [hep-ph/9902341]\,.

\bibitem{KNMW}
Z.~Kunszt, P.~Nason, G.~Marchesini and B.~R.~Webber, Z Physics at LEP 1, preprint
CERN 89-08, vol. 1, pp.~373-453.

\bibitem{ERT}
R.~K.~Ellis, D.~A.~Ross and A.~E.~Terrano,
Nucl.\ Phys.\  {\bf B178} (1981) 421.

\bibitem{CTTW}
S.~Catani, L.~Trentadue, G.~Turnock and B.~R.~Webber,
Nucl.\ Phys.\  {\bf B407} (1993) 3.

\bibitem{CW}
S.~Catani and B.~R.~Webber,
Phys.\ Lett.\  {\bf B427} (1998) 377 [hep-ph/9801350].

\bibitem{DW}
Y.~L.~Dokshitzer and B.~R.~Webber,
Phys.\ Lett.\  {\bf B404} (1997) 321 [hep-ph/9704298].

\bibitem{NS}
P.~Nason and M.~H.~Seymour,
Nucl.\ Phys.\  {\bf B454} (1995) 291 [hep-ph/9506317].

\bibitem{Nonloc}
F.~R.~Ore and G.~Sterman,
Nucl.\ Phys.\  {\bf B165} (1980) 93;
\\
G.~P.~Korchemsky, G.~Oderda and G.~Sterman, in Proceedings of the 5th
International Workshop on Deep Inelastic Scattering and QCD (DIS 97), Chicago,
IL, 14-18 Apr 1997
[hep-ph/9708346].

\bibitem{CS}
N.~A.~Sveshnikov and F.~V.~Tkachev,
Phys.\ Lett.\  {\bf B382} (1996) 403 [hep-ph/9512370];
\\
P.~S.~Cherzor and N.~A.~Sveshnikov,
hep-ph/9710349

\bibitem{G-M}
E.~Gardi, Talk at the 35th Rencontres de Moriond, Les Arcs, France, 18-25 Mar
2000 [http://moriond.in2p3.fr/QCD00/transparencies/4\_wednesday/pm/gardi/].

\bibitem{G}
E.~Gardi,
JHEP {\bf 0004} (2000) 030 [hep-ph/0003179].

\end{thebibliography}
\end{document}